\documentclass[12pt]{paper}

\usepackage{amsmath}
\usepackage{amsfonts}
\usepackage{graphicx}
\usepackage{todonotes}

\usepackage{color}

\newcommand{\h}{h_{\alpha\beta}}
\renewcommand{\d}{\mathrm{d}}

  \usepackage{natbib}
\title{Lost Horizon? -- Modeling Black Holes in String Theory} 
\author{Nick Huggett \& Keizo Matsubara\thanks{Some of this work was performed under a collaborative agreement between the University of Illinois at Chicago and the University of Geneva and made possible by grant numbers 56314 and 61387 from the John Templeton Foundation. Its contents are solely the responsibility of the authors and do not necessarily represent the official views of the John Templeton Foundation. We also thank John Dougherty for considerable assistance at the start of this work, and an anonymous referee.}}

\date{}
\begin{document}
\maketitle

\noindent Department of Philosophy\\ 
University of Illinois at Chicago\\
601 South Morgan Street\\
Chicago, Illinois 60607, USA\\
{\tt huggett@uic.edu}\\ 

\noindent Department of Philosophy\\
Uppsala University\\ 
Box 627, SE-751 26 Uppsala, Sweden \\
{\tt keizo.matsubara@filosofi.uu.se} 

\begin{abstract}
The modeling of black holes is an important desideratum for any quantum theory of gravity. Not only is a classical black hole metric sought, but also agreement with the laws of black hole thermodynamics. In this paper, we describe how these goals are obtained in string theory. We review black hole thermodynamics, and then explicate the general stringy derivation of classical spacetimes, the construction of a simple black hole solution, and the derivation of its entropy. With that in hand, we address some important philosophical and conceptual questions: the confirmatory value of the derivation, the bearing of the model on recent discussions of the so-called `information paradox', and the implications of the model for the nature of space.
\end{abstract}

\newpage

\section{Introduction}
In their article on singularities and black holes in the Stanford Encyclopedia of Philosophy, Peter Bokulich and Erik Curiel  raise a series of important philosophical questions regarding black holes, including the following:

\begin{quote}
When matter forms a black hole, it is transformed into a purely gravitational entity. When a black hole evaporates, spacetime curvature is transformed into ordinary matter. Thus black holes appear to be crucial for our understanding of the relationship between matter and spacetime, and so provide an important arena for investigating the ontology of spacetime, of material systems, and of the relations between them. \cite{CurBok:12}
\end{quote}

This paper develops this insight to investigate the natures and relations of spacetime and matter in quantum gravity, specifically in string theory. Part of the paper will therefore be devoted to explicating the general status of spacetime in string theory (\S\ref{sec:strings}), and especially its emergence, and fungibility with matter; and to outlining a well-studied example of a string theoretic black hole (\S\ref{sec:stringBH}).

As Bokulich and Curiel note, of particular significance in such an investigation is the phenomenon of black hole thermodynamics (BHT) and Hawking radiation. This topic has been widely discussed by philosophers of physics as well as physicists, so we will just give a brief review (\S\ref{sec:BHT}). It is important to note that while much theoretical work motivates the results of BHT, there is no direct empirical confirmation of these results. The only experimental evidence comes from work on analogue systems, whose significance remains controversial.\footnote{For instance, see \cite{DarThe:17}, and \cite{Cro:19}.}  The result that black holes radiate is, nonetheless, generally trusted since the derivations rely on well tested theories, applied in regimes where we should be able to trust the derived conclusions. 
Once we have described a stringy black hole, and the agreement of Boltzmann and Bekenstein-Hawking entropies, we will discuss the epistemic significance of this result (\S\ref{sec:SBHentropy}). 

However, our focus is on the ontological aspects of black holes, and we will argue (\S\ref{sec:info}) that the issues that arise under the heading of the `information paradox', such as the unitarity of black hole evaporation, and the possibility of `firewalls' or `fuzzballs', suggest insights into the nature of spacetime in the interior of a stringy black hole. Then we will turn to some more general lessons about the relation between space and matter, to be drawn from our discussion (\S\ref{sec:lessons}).

\section{Black hole thermodynamics}\label{sec:BHT}

Assuming some familiarity concerning the topic of BHT, this review will be brief.\footnote{That black holes have entropy was originally claimed in \cite{Bek:73}. After \cite{Haw:75} showed that black holes radiate, BHT was taken much more seriously. Philosophical work on BHT include \cite{BelEar:99}, \cite{Wal:18,Wal:19,Wal:20}, and \cite{Wut:17}. Reviews by physicists include \cite{SusLin:05}, \cite{Mat:09}, \cite{Har:16}, and \cite{Pol:17}.} Starting with classical general relativity (GR) without quantum effects, it was observed that black holes seemed to violate the second law of thermodynamics: dropping things into a black hole could seemingly destroy entropy. To avoid this conclusion,  invoking quantum considerations \cite{Bek:73} proposed an entropy proportional to the area of the horizon. Keeping all physical constants explicit, the formula for the entropy is as follows:

\begin{equation}
\label{eq:SBH}
S_{BH} =\frac{k_B c^3 A}{4 \hbar G}= \frac{k_B A}{4 \ell_p^2} 
\end{equation}
using that the Planck length $\ell_p = \sqrt{\frac{G\hbar}{c^3}}$, to display more explicitly how the area of the black hole is divided into Planck length squared areas.

There is an incompatibility between the  quantum Bekenstein entropy, the Boltzmannian understanding of thermodynamics, and the  classical ``no hair theorem''. On the one hand the entropy should be attributed to the (logarithm of) the number of black hole microstates. On the other, in classical GR the state of a black hole is \emph{completely} characterized by its mass, charge and angular momentum -- ``black holes have no hair''.  Perhaps unsurprisingly, classical black holes simply don't have the microstates necessary to understand the Bekenstein entropy in Boltzmannian terms.

Perhaps black holes have a different kind of entropy; indeed, \cite{Haw:75, Haw:76} essentially propose that they demonstrate the existence of non-Boltzmannian entropy in quantum mechanics.  However, physicists working in the different approaches to quantum gravity generally aim to provide a description of the quantum microphysics of black holes. If such an account can be given, the Boltzmann picture ``assures'' us that some form of the second law holds  even when systems include black holes: by state space volume considerations, most states at lower entropy evolve to states of higher entropy. 

Now, if black holes are properly thermodynamical, then there should also be a temperature associated with them, and they should seek thermal equilibrium with their environment. Of course, Hawking radiation provides a realization of just this. However, it also allows for `information loss': in spacetimes containing black holes, pure quantum states can evolve into mixed states. However, quantum physics is unitary, and it is a theorem that unitarity prohibits the evolution from a pure state to mixed one. Moreover, such an evolution amounts to a failure of \emph{backwards} determinism: one cannot retrodict an earlier pure state from a mixed state. This surprising conclusion led to much debate on the so-called ``black hole information paradox'' (or ``problem'').

One response is `black hole complementarity'\footnote{See \cite{SusTho:93}.}, whose central idea is that external observers never see matter entering the horizon, because of the infinite red shift, and instead observe it radiating back in an unproblematic way. Observers that do cross the horizon of course do see matter entering the black hole, but are shielded from observing any inconsistency (specifically, violations of the quantum no cloning theorem) because they fall into the singularity too quickly.\footnote{ See \cite{BelEar:99} and \cite{van:04} for philosophical discussions.} In response, \cite{AlmMar:13} aimed to show that three claims assumed by complementarity are inconsistent.
 Quote:

\begin{enumerate}
\item \textbf{Unitarity}: Hawking radiation is in a pure state. 
\item \textbf{Semi-classical gravity}: The information carried by the radiation is emitted near the horizon, with low energy effective field theory valid beyond some distance from the horizon.
\item \textbf{No drama}: The infalling observer encounters nothing special at the horizon.
\end{enumerate}

\noindent To avoid contradiction, Almheiri et al. deny (3), proposing that an infalling observer does not pass the horizon as expected classically, but instead is destroyed by a `firewall'; which certainly would be drama! In \S\ref{sec:FW} we will return briefly to firewalls; while in \S\ref{sec:fuzz} we shall see another view, which also rejects 3.

\section{Spacetime in string theory}\label{sec:strings}

Before continuing with these issues, we need to outline sufficient string theory to understand how stringy black holes arise: the origin of both spacetime (\S\ref{sec:STSP}) and matter (\S\ref{sec:STM}) in the theory. 

\subsection{Spacetime in string theory: fungibility of geometry and matter}\label{sec:STSP}

\cite{HugVis:15} explained the derivation of the Einstein Field Equation (EFE) -- the `emergence' of GR -- in string theory. Since this story is central to the points of this paper we must review it, but with emphasis on the conceptual picture, and without the technical details found in that paper.\footnote{Or in the sources from which it is drawn, e.g.\ \cite{Pol:03}. See \cite{Vis:19} or \cite{Nic:20} for longer philosophical analyses.}  The essential ontological innovation is the blurring of the space-matter distinction.

\begin{figure}[htbp] 
  \centering
   \includegraphics[width=3in]{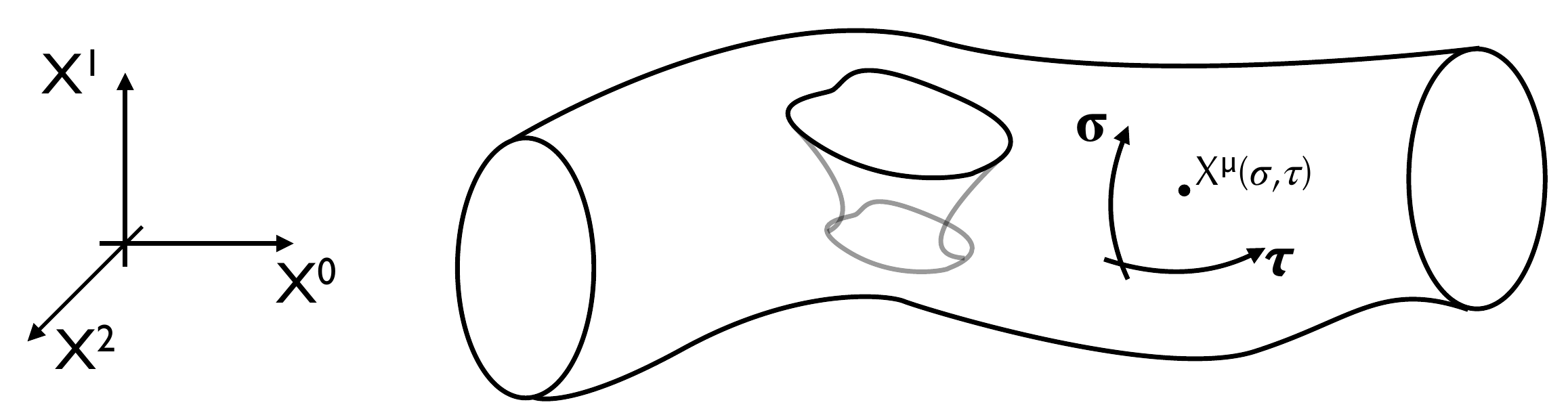} 
   \caption{A closed string in spacetime. The trajectory is described by an embedding function from worldsheet coordinates to spacetime coordinates: $(\sigma,\tau)\to X^\mu$.}
      \label{fig:NG}
\end{figure}

The  starting point for classical string theory is the Nambu-Goto action, which tells us to extremize the worldsheet spacetime area of a string in a $d$-dimensional Minkowski background (figure \ref{fig:NG}). So doing leads to a relativistic wave equation, with either Neumann (momentum conserving) or Dirichlet (position conserving) boundary conditions at the end points.

However, the Nambu-Goto formulation is infelicitous for quantization, so one shifts to the classically equivalent Polyakov action (see (\ref{eq:poly}) below). So doing introduces an `auxiliary' Lorentzian metric $\h$ on the string worldsheet, \emph{distinct} from the metric `induced' on the world sheet by the Minkowski metric of background spacetime. (The subscripts range over the two coordinates $\sigma$ and $\tau$ on the worldsheet.) Importantly, the action has `Weyl symmetry' with respect to $\h$: $\h\to e^{\Omega(\sigma,\tau)}\h$ for any smooth real function $\Omega(\sigma,\tau)$. Thus there is no physical significance to the auxiliary metric beyond the causal structure it ascribes to the string, which must agree with that of the background spacetime in order to minimize the action. 

On canonical quantization, the classical wave solutions become quanta \emph{on} the string, in the way familiar from quantum field theory (QFT), which when grouped into states of equal energy form representations of $SO(1,d-1)$, just like relativistic particles in $d$-dimensional spacetime. Hence particles are reinterpreted as strings in the appropriate representations, with rest mass associated with the vibrational energy of the string -- at length scales at which the string is indistinguishable from a point. By this mechanism string theory promises to unify the different fundamental particles: they are nothing but different modes of a single underlying object, the string, and hence fungible if the state of the string changes. In particular, the spectrum of the closed bosonic string contains the massless spin-2 representation that characterizes the graviton, the quantum of the metric field; these modes/particles are therefore in particular fungible with those of other fields. That said, several points should be made.

First, we are yet to identify \emph{quanta} of the corresponding quantum fields as strings, since creation and annihilation of quanta requires creation and annihilation of strings, about which nothing has yet been said. Modes on a string can be created and annihilated, but that does not change the number of strings, just the kind of particle that a string represents. Second, massless spin-2 fields lead almost inevitably to GR: classically see \citet[\S18.1]{MisTho:73}, while \cite{Sal:18} reviews the situation in QFT. So if this mode of the string truly is a quantum of the gravitational field, we need to verify that it relates dynamically to other fields in the appropriate way -- through the EFE. Third, the bosonic string is incapable of reproducing the mass spectrum of the standard model; again, more structure must be added. All three points will be developed later.

Progressing further requires shifting to a path integral approach, in which each path contributes an amplitude equal to the exponential of its action, or rather $e^{iS}$. Wick rotating the worldsheet coordinates $\tau\to i\tau$ to give the auxiliary metric $\h$ a Euclidean signature, the Polyakov path integral is given by \citet[\S3.2]{Pol:03}:

\begin{equation}
\label{eq:poly}
\int_{paths}DXDh\ \exp\big\{\frac{-1}{4\pi\alpha'}\int_M\d\sigma\d\tau\ h^{1/2}h^{\alpha\beta}g_{\mu\nu}\partial_\alpha X^\mu\partial_\beta X^\nu\big\},
\end{equation}
where the `Regge slope' $\alpha'$ is the characteristic string length squared, $M$ is a specified worldsheet, and (for now) $g_{\mu\nu}=\eta_{\mu\nu}$, a background Minkowski metric. The path integral is taken over all embeddings $X^\mu$ and all auxiliary metrics $\h$.

The path integral involves a sum over all topologically distinct worldsheets: for the closed string, tori of all possible genera, with $N$ open holes representing in/out strings at temporal infinity. The topological holes in the tori are produced by strings splitting/joining: for instance, figure \ref{fig:NG} is a simple torus with $N=2$, representing a single incoming closed string splitting into two strings, which then recombine into a single outgoing string. The tori therefore represent a perturbative sum of Feynman diagrams, in analogy with those for QFT (indeed under the identification of quanta with string modes, QFT diagrams are understood as approximations to stringy diagrams). 

Therefore they assume the existence of a theory in which strings can be created and annihilated, or at least  of a theory  to which Fock-like string states are a reasonable approximation (in some sector). (In)famously, this  latter theory -- `M-theory' -- is not known, and so string theory as we are discussing it is inherently perturbative.\footnote{A bosonic string field theory, with a 3-point interaction exists (e.g., \cite{Tay:09}), but is  not viewed as a candidate fundamental string theory.} However, once one accepts this perturbative understanding then the identification of strings with the quanta of QFT is complete: any field state (in the Fock representation, a superposition of different numbers of quanta) is fundamentally a state of many strings (a superposition of different numbers of strings, each in the mode corresponding to the quantum of the field). Thus all fields are unified, composed of strings, differing only in their modes, and fungible if the strings change mode. We now have all the conceptual ingredients needed to understand the origin of GR in string theory. 

(i) First, GR allows for curved background spacetime metrics, not just Minkowski spacetime. In QFT, classical fields are represented by `coherent states' of field quanta. Such states can be defined in various ways (see \cite[\S8.2-3]{Dun:12}), but two conceptions are salient: first, they are maximally classical, simultaneously minimizing the uncertainty in the canonical variables; second, they are collective states, involving a superposition of every number of field quantum (and so are not finite superpositions). But if a classical field is described by a coherent state of quanta, then according to the identification of quanta with strings, a classical field should correspond to a suitable collective superposition of strings, each excited into the same mode. The story will be the same for any classical field, including a metric field comprised of stringy gravitons. 

One can check this identification, by inserting classical fields into the Polyakov action, and comparing the effect on scattering amplitudes with that of scattering in the presence of the corresponding collective string states. For instance, one might take $g_{\mu\nu}$ to be a general spacetime metric rather than Minkowski, and compare it with scattering in a background of a suitable coherent state of stringy gravitons. The results are exactly the same: these are equivalent descriptions.\footnote{ \citet[\S3.4.1]{GreSch:87} give the following demonstration of their equivalence: a coherent state of strings, each in a massless spin-2 state, introduces a term $\gamma_{\mu\nu}$ in the path integral (\ref{eq:poly}) which adds to the Minkowski metric to produce $\eta_{\mu\nu}\to g_{\mu\nu}=\eta_{\mu\nu} + \gamma_{\mu\nu}$. Since the path integral determines all physical quantities in a quantum theory, we have fully equivalent theories whether we introduce the curved metric as a classical field or as a graviton state.} Note that the classical fields are called `background' fields, but in the sense that they describe a fully stringy background, not because they are added to the theory from the outside. 

(ii) Second, a path integral like (\ref{eq:poly}) with a general curved metric is known as a `non-linear sigma model'; broadly, it describes a field $X^\mu$ living on a 2-dimensional spacetime (the string worldsheet) with \emph{variable} interaction $g_{\mu\nu}(X^\mu)$. The crucial result for our purposes is that this quantum theory will only retain the Weyl invariance of the classical action -- as it must do in order to avoid a pathological `anomaly' -- if the background metric $g_{\mu\nu}$ and any other background fields satisfy the EFE (to lowest order in $\alpha'$ ). For (\ref{eq:poly}), in which there is only a background metric field, the result is the free field equation $R_{\mu\nu}=0$; in general, with additional background fields, the \emph{full} non-linear equation is entailed.\footnote{It is worth stressing that the expansion is in $\alpha'$, so that the approximation is prima facie valid when the radius of spacetime curvature is small compared to the string length: say, compared to the Planck length -- far beyond the regime of linear gravity. We will, however, see that it seems to break down in a `fuzzball', even for moderate curvature.} Of course, from our previous discussion, we recognize that the metric (and other) fields are in fact nothing but collective string states.

To summarize: avoiding a Weyl anomaly requires that background fields, including the metric, satisfy the EFE to lowest order in perturbation theory. Physically however, the background does not comprise classical fields in a classical spacetime: rather strings in appropriate modes form coherent states of effective QFTs, which in turn form effective classical fields. So ultimately the Weyl anomaly is a constraint on multi-string states, and the ontology of fields is one of strings only. But since the quanta of different fields, including the metric, are nothing but different string modes, they are fungible, so that gravity is on the same footing with any other force.\footnote{True, the full metric contains Minkowski and stringy parts: $g_{\mu\nu}=\eta_{\mu\nu} + \gamma_{\mu\nu}$. But the conclusion that $\eta_{\mu\nu}$ is a non-stringy classical background can be resisted: \cite{wit96,Mat:13, Hug:15,Mot:12}. See \cite{Rea:19} for more on fungibility.}

\subsection{Supergravity: stringy fermions, gauge fields, and $p$-branes}\label{sec:STM}

Since the world contains fermions one must extend string theory: as bosons arise from spatial modes, so fermions arise from vibrations in `anti-commuting directions'. A full discussion is well beyond the scope of this paper so we will only sketch points necessary for our string theoretic black hole model. The most important point is that the recovery of GR from string theory just described applies \textit{mutatis mutandis} to superstring theory.

In very general terms, `supersymmetric' (SUSY) string theory is developed as for the bosonic string. First introduce an action that adds fermionic degrees of freedom $\psi^\mu(\sigma,\tau,)$ (a Majorana spinor) to the bosonic ones $X^\mu(\sigma,\tau,)$:

\begin{equation}
\label{eq:SUSY}
\int_{paths}DXDh\ \exp\big\{\frac{-1}{4\pi\alpha'}\int_M\d\sigma\d\tau\ h^{1/2}h^{\alpha\beta}g_{\mu\nu}(\partial_\alpha X^\mu\partial_\beta X^\nu - i\psi^{\dagger\mu}\rho_\alpha\partial_\beta\psi^\nu)\big\},
\end{equation}
where $\rho^\alpha$ are worldsheet Dirac matrices.  \citet[\S4.1]{GreSch:87} discusses this action, and shows that it possesses classical supersymmetry. There are new endpoint boundary conditions for the fermionic degrees of freedom -- not Neumann and Dirichlet, but Ramond or Neveu-Schwarz -- and correspondingly new modes. When one canonically quantizes, one's choice of boundary conditions produces a particular spectrum of bosons and fermions. Because of  the underlying SUSY these are paired (in addition to \citet[\S4.2]{GreSch:87}, \citet[chapters 14-6]{Zwi:04} contains an approachable introduction): each mode is fungible with its `superpartner', under a symmetry of the theory. 

Proceeding exactly as before, the bosonic modes correspond to field quanta, but now of gauge fields. Coherent states of strings in the same mode thus have effective descriptions as classical gauge potentials, $A_\mu, A_{\mu\nu}, A_{\lambda\mu\nu}$, and so on. And of course to avoid the Weyl anomaly, with the metric these mutually satisfy the appropriate EFE, and hence because of their supersymmetry form models of classical `supergravity'. 

The question arises of the sources of these fields. $(n-1)$-dimensional bodies can couple `electrically' to an $n$-form gauge field. For instance, a 0-dimensional point body couples as $A_\mu\frac{\d x^\mu(\tau)}{\d\tau}$: the dimension of the body determines whether it has enough indices to `eat' the field indices. Similarly,  $d-n-3$ dimensional objects couple `magnetically' (since they have to `eat' the indices on the field's Hodge dual). So the presence of gauge fields speaks for the presence of charged multidimensional objects, known as `$p$-branes'.  These are typically thought of in terms of stable `solitonic' multi-string states, but they also ground Dirichlet boundary conditions in string theory: if the end of a string is constrained to move within a $p$-brane, then it is fixed with respect to the remaining $d-1-p$ spatial dimensions. A $p$-brane to which open strings can attach is thus known as a D$p$-brane. (See \citet[\S2.2]{de20} for the conceptual development of this idea, and the important role of \citet{pol:95}.)  

Pulling this together, one of the choices of boundary condition leads to `type IIB' superstring theory, which contains a 2-form gauge field $B_{\mu\nu}$. So, for example in 10 spacetime dimensions, D1-branes couple electrically and D5-branes magnetically to $B_{\mu\nu}$, and so may be present in a supergravity limit of type IIB superstring theory. In our model, a construction of these branes produces the black hole. 

\section{A stringy black hole}\label{sec:stringBH}

In this section we sketch a realization of the preceding ideas, a stringy black hole, which will be the basis for the following discussion. Our model is physically unrealistic (at least for our universe), but it is simple yet exhibits the principles behind more realistic examples.\footnote{ Hence it is popular in pedagogical presentations, e.g. \citet{DasMat:00} and \citet[chapter 22]{Zwi:04}.) The origin of this type of construction to show that the entropy agrees with the Bekenstein-Hawking formula (\ref{eq:SBH}) is \cite{StrVaf:96}, but the specific approach discussed was proposed in \citet{ca:96}. It is studied in historical and conceptual depth in \citet{de20}: inter alia, this paper describes the state of string theory before and after 1996, the technical and conceptual details of the calculation, and its subtle logic (drawing on a number of approximations and correspondences) and limitations. We strongly recommend it for a full treatment beyond the sketch given for our purposes (see also \citet{DonDe-:20}.} With this model in hand we can turn to our philosophical themes: the implications of stringy black holes for the matter-space relation and BHT.

We work in type IIB theory with its D1- and D5-branes, and suppose a background spacetime topology of $R^5\times S^1\times T^4$ with coordinates $(x_0, \dots x_4, x_5,\\ x_6, \dots x_9)$, respectively. We are interested in the black hole appearing in the 5-dimensional spacetime described by $(x_0, \dots x_4)$ with topology $R^5$, and stipulate that the remaining compact dimensions are `small'. However, the circumference $C$ of the circular $S^1$ $x_5$ dimension is much larger than that of the toroidal $T^4$ $(x_6, \dots x_9)$ dimensions. The effect of this stipulation is that the minimum wavelength on the torus is much shorter than on the circle, so that the energy cost of excitations on the torus is much greater, and effectively any momentum in the compactified dimensions will be on the circle. Thus if $N$ is the wavenumber on $S^1$, then $C/N$ is the wavelength, and the internal momentum of the system is $P=hN/C$. 

   \begin{figure}[htbp] 
     \centering
     \includegraphics[width=3.5in]{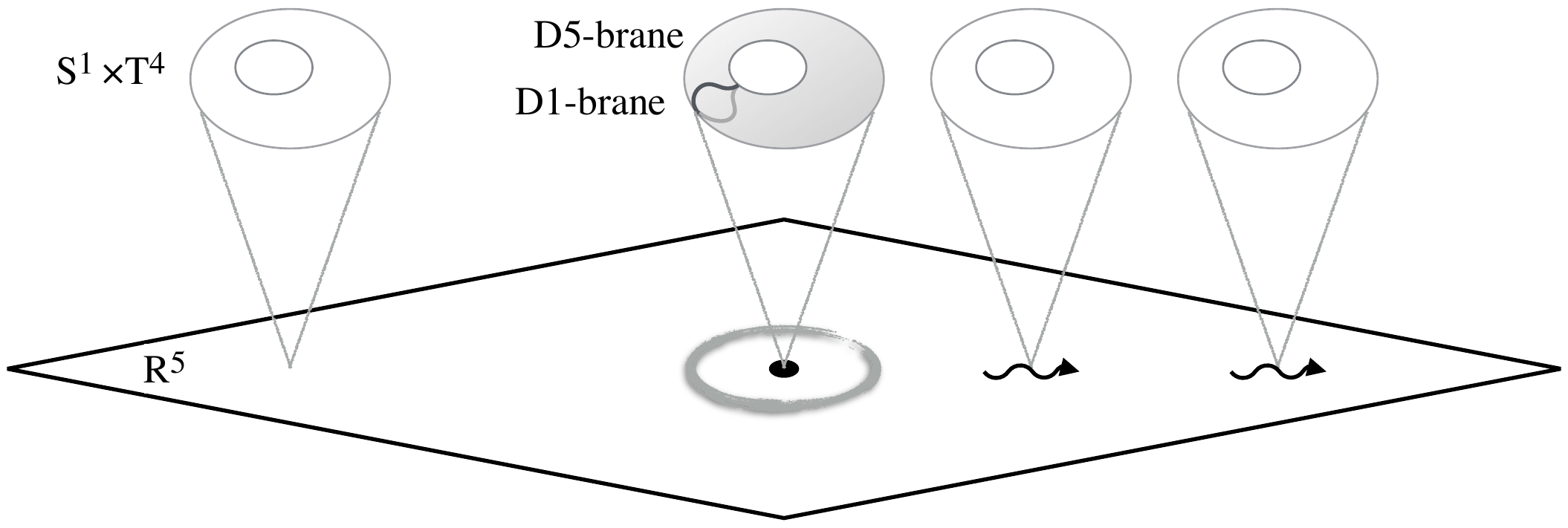} 
  \caption{A stringy black hole: the background spacetime has a topology $R^5\times S^1\times T^4$ -- time is not shown, and of space $S^1$, two dimensions of $R^4$ and one of $T^4$ are pictured. At a point of $R^4$ are located D1-branes around $S^1$ and D5-branes around $S^1\times T^4$. If the string interaction is `turned on', a spatial horizon forms around the branes in $R^5$, and energy is radiated.}
        \label{fig:BH}
  \end{figure}

At the origin of the uncompactified space, $(x_1, \dots x_4)=(0,0,0,0)$, are located (a) $Q_1$ D1-branes wrapped around $S^1$, (b) $Q_5$ D5-branes wrapped around $S^1\times T^4$, and (c) momentum $P$ (in the $x_5$ direction, as discussed); see figure \ref{fig:BH}.  As we saw, the D$p$-branes couple to the $B_{\mu\nu}$ gauge field of the theory (whose stringy nature we again emphasize), while $P$ is a source for the metric field $g_{\mu\nu}$ (likewise). That is to say, assuming that strings interact, $Q_1$, $Q_5$, and $P$ act as charges for these background fields; then, because the EFE holds for them (to avoid the Weyl anomaly), the spacetime geometry in which the system lives can be computed. One then applies the technique of `dimensional reduction' based on the work of Kaluza and Klein (see \cite{Kar:12}), to determine the projection of higher dimensional physics into the large dimensions that we directly observe. Gauge fields project into gauge fields, but so does the metric: from the point of view of the large dimensions, the geometry of the compact dimensions acts as if there was a new gauge field -- the basis of the Kaluza-Klein scheme to `geometrize' gauge fields. The upshot is that the $R^5$ supergravity description of the solution is a Reissner-Nordstr\"om black hole with three point charges $Q_1$, $Q_5$, and $P$, and mass equal to its internal energy, located at the origin as shown in figure \ref{fig:BH}.

The point of constructing such models was to calculate their Boltzmann entropy and compare it with $S_{BH}$ (\ref{eq:SBH}) for this supergravity black hole; they indeed take the same value, $S\sim\sqrt{Q_1Q_2P}$. (Undermining the idea that black holes might have non-Boltzmannian entropy.) The details and justification of the calculation are beyond the scope of this paper, but we will emphasize an important point.\footnote{Three other comments: (1) The calculation does not appeal to perturbative string theory laid out earlier. (2) For details we refer the reader to our other references, and especially to \citet{de20} for the role of BPS states (\S3.1), and the complex relations between the free and interacting pictures (\S3.3) justifying the result. (3) We thank a referee for patiently clarifying the significance of the following point.} The Boltzmann entropy is associated with the microstates of the brane assemblage, and the techniques for counting these only apply to \emph{non-interacting} string theory: to an `ideal D$p$-brane gas'. `Turning off' the string interaction is physically meaningful, since it is a dynamical parameter of the theory, but the result is that stringy gravitons no longer interact with stringy matter, and the gravitational force is turned off: hence no black hole forms. In this sense the brane system is not a \emph{literal} description of the black hole source, but merely \emph{corresponds} in significant physical regards to the interior of the black hole.

The nature of this correspondence is (also) beyond the scope of this paper, but of course important for fully understanding the nature of a stringy black hole, beyond the perturbative description of spacetime which we have described.  For a careful  analysis we refer the reader to \citet{DonDe-:20}, in which it is argued (\S3) that the relation is not one of strict equivalence, but (\S4) of ontological emergence. (We come at the question of the string state of the black hole interior from a different point of view below.) 

However, for current purposes, the important point is that the validity of the calculation turns on the argument that the number of microstates is independent of the strength of the string interaction: that it is the same in both the supergravity and D$p$-brane gas systems. Of central (but not unique) importance here is that the system is in a Bogomol'nyi-Prasad-Sommerfield (BPS) state of superstrings (\citet[\S3.2]{DasMat:00} gives a simple illustration). These arise in SUSY because of the special symmetries, but have the feature that varying interaction terms does not cause  splitting of energy levels, so that the number of states indeed remains constant.\footnote{\label{fn:ad}An earlier program due to Susskind, on which he reflects in (\citeyear{Sus:06}), approached the same problem by adiabaticity; that slowly lowering the string interaction to zero would not change the state counting. This method is more general, allowing the Boltzmann entropy to be calculated for a range of realistic, non-extremal black holes, but is less reliable because it doesn't have the BPS guarantee that the density of states is constant.} However, a selection rule means that they are energetically stable, implying that the black hole in the effective supergravity model is `extremal', unable to Hawking radiate any further, though not completely evaporated away.

However, as \cite{Wad:01} explains, one can treat a weakly interacting system of D$p$-branes corresponding to a near-extremal black hole perturbatively to verify that the Boltzmann and Bekenstein-Hawking entropies agree as well. Most significant for our discussion, there is a channel by which branes can radiate gravitons into $R^5$ (\citet[\S6.2]{DasMat:00}). That is, if $\Phi^I$ ($I=7,8,9,10$) represents a quantum of D1-brane vibration in the $T^4$ directions, and $h_{IJ}$ a graviton polarized in the $T^4$ dimensions propagating in $R^5$, then the following interaction exists:

     \begin{equation}
\label{eq:rad}
\centering
     \includegraphics[width=.5in]{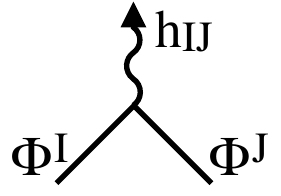} 
\end{equation}
That is, the model has a mechanism for the black hole to radiate mass away; moreover, the quantum mechanical -- unitary -- energy cross-section of this radiation agrees with that computed semi-classically for Hawking radiation. Granted, these results are derived in the brane system rather than the supergravity black hole, but arguably the correspondence continues to hold, and they indicate that stringy black holes indeed posses Bekenstein entropy and Hawking radiate unitarily (figure \ref{fig:BH}). Moreover, such an interaction provides a specific instance of how the fungibility of string modes, especially those of matter and geometry, play out in dynamical processes.

\section{String theory and black hole thermodynamics}\label{sec:info}

We have described, with a focus on their conceptual significance, black hole thermodynamics (\S\ref{sec:BHT}),  the standard string theoretic interpretation of spacetime (\S\ref{sec:strings}), and stringy black holes (\S\ref{sec:stringBH}). Now we turn to a fuller investigation of their philosophical consequences. In this section we will focus on BHT. We consider a different black hole model in whose interior, spacetime, in the standard interpretation, might break down; and some consequences of this for the information loss paradox. In \S\ref{sec:lessons} we will draw more general conclusions for the nature of spacetime in stringy black holes.

\subsection{Significance of the entropy agreement}\label{sec:SBHentropy}

But first, while the focus of this paper is the ontology of stringy black holes, some brief comments on their epistemic import are in order. Especially, what is the confirmatory value for string theory of the equality of Boltzmann and Bekenstein-Hawking entropies?  Why are such results considered important, given that there is no direct empirical confirmation of BHT? We will make four points.\footnote{See \cite{Wal:18,Wal:19}, \cite{DonDe-:20} and \citet{de20} for more detailed discussions.}

First, there are nonetheless reasons to trust BHT, especially the consilience of many routes to their derivation, across multiple contexts; and with the general framework of thermodynamics, beyond gravitational physics. (And perhaps analogue experiment.)

Second, Bekenstein's discovery was `surprising', which might make it seem a particularly strong piece of evidence. However, the surprise is not of the evidentially relevant kind. We must distinguish the \emph{anticipation} that $P$ is true from the \emph{probability} that $P$ is true conditional on our background beliefs. For the confirmation of string theory our background beliefs include semi-classical GR, the theory of quanta propagating in curved spacetimes; as we saw, recovering this theory is already part of the support for string theory. But as Hawking showed, BHT is a consequence of semi-classical GR, and so not independent evidence; the surprise was only that of discovering an unknown logical consequence.

Third, indeed, since string theory is believed to have QFT and GR as effective limits,\footnote{See \citet[\S3.3]{de20} for a detailed discussion of the role of this assumption in the entropy calculation.} it `must' entail BHT. Thus it is better to take the successful entropy derivation for stringy black holes as a consistency check rather than new evidence. For instance, \citet[p1]{HorMal:96} seems to express this attitude. However, success is non-trivial: if one could  show the failure of only one model to be consistent with the results of BHT then this would be highly problematic for string theory.

Finally, the derivation provides a concrete (if non-literal) account of the microstates of a black hole, showing the power of the string theoretic principles assumed in modelling it: i.e., the assumptions of the previous section and subsection. This account goes beyond general semi-classical GR, and so does receive confirmation from the derivation of BHT. The nature of this confirmation (and of the assumptions) is analyzed in far greater detail in \citet{de20}, which also emphasizes the `heuristic' value of the successful modeling principles. We would put the point this way:  success provided two other kinds of support for string theory. In the first place, the derivation of a Boltzmann entropy is a kind of explanation of the Bekenstein-Hawking entropy (and more speculatively of Hawking radiation); and so provides whatever theoretical support successful explanations give. Second, the constructions licensed by these assumptions allow the application of string theory to further physical situations, in spacetime and particle physics, increasing its fruitfulness, an important non-empirical virtue of theories.

\subsection{Reflections on the black hole information paradox}

As a result of Strominger and Vafa's successful calculation of stringy black hole entropy, most string theorists were convinced black holes evaporate unitarily, and that Hawking's proposed information loss would be resolved in a full theory of quantum gravity (\citet{de20}). On the other hand, it has been argued that there would, in any case, be no `paradox' in a failure of unitarity in black hole evaporation. We address these points through the ``firewall" and ``fuzzball'' proposals, to illuminate how unitarity might be obtained in string theory, and why, after all, there would be a `paradox' for string theory if it were not. At the heart of this discussion is the important idea that there may be no spacetime at all in the interior of a black hole.

\subsubsection{On Maudlin's recent critique of the information paradox}

\cite{Mau:17} recently argued that the whole idea of a paradox is due to a simple mistake. He first observes -- reiterating \cite{Wal:94} -- that the spacelike surfaces after the evaporation of the black hole are not Cauchy-surfaces: causal curves from the past can end up in the singularity, and fail to reach post-evaporation hypersurfaces. But  it is only for an evolution of a pure state from one Cauchy-surface to another that the rules of quantum mechanics imply that the state must remain pure. The more novel suggestion made by Maudlin is that therefore the final mixed state does \emph{not} require that the evolution is not unitary. To make this point he uses a slightly unconventional foliation of spacetime, where some of the Cauchy-surfaces are disconnected; with respect to this foliation the full evolution is unitary. 

While we do not dispute these technical claims, we believe that physicists working on the black hole information paradox generally are aware of Wald's argument, and won't be moved by Maudlin's conclusions. In particular, his description of the situation presupposes the classical, GR description of spacetime everywhere, but this cannot be taken for granted in a theory of quantum gravity.\footnote{\cite{HugWut:13a} explores spacetime emergence.}
 It is true that he offers a way to reconcile classical spacetime with unitarity, but it only diagnoses the loss of information rather than removing it. Many working in the field expect a full quantum gravity description of the formation and evaporation of black holes not to involve any singularities or loss of causality, and yet remain unitary. The `paradox' is that this does not occur in the combination of the two theories, GR and QFT, that are presumably low energy limits of the fundamental theory.\footnote{In addition, see \cite{Wal:20} for a convincing demonstration that there are forms of the paradox that resist Maudlin's analysis.}  From this point of view the question remains of how to reconcile unitarity with external observers who do not encounter information loss.

\subsubsection{Firewalls}\label{sec:FW}

In \S\ref{sec:BHT} we described the `AMPS' argument  (\cite{AlmMar:13})  that the premises assumed by black hole complementarity were not consistent. A number of different responses have been formulated (see \cite{Pol:17} and \cite{Har:16}). One is to accept `drama' at the horizon, or even the absence of a horizon in the first place. Objects -- and observers -- never really pass the horizon, instead they are thermalized before they reach it; there only is the exterior description, no complementary description according to an infalling observer.  In this case, it has been suggested that there is no classical spacetime interior either:

\begin{quotation}
Finally, since we are thinking that spacetime is emergent, we might try the slogan that it is not that the firewall appears, but that the interior spacetime fails to emerge. But to claim this we would need a better understanding of emergent spacetime.\\
(\citet[31]{Pol:17}.)\footnote{See \cite{Sus:12,Sus:12a,Sus:12b} for further discussions.}
\end{quotation}

From Maudlin's point of view this suggestion might seem irrelevant, since he does not accept the premises that motivated the introduction of the firewall in the first place. But from the point of view of a quantum theory of gravity in which spacetime is emergent, such a view certainly makes sense; the interior could be described by fundamental degrees of freedom that do not have a classical spacetime description. And if the black hole interior is eliminated, then the disconnected parts of the Cauchy-slices to which Maudlin appeals for unitarity will also be eliminated. As we noted, we believe that Maudlin's reasoning takes a fundamental spacetime for granted, when this is often denied by those in the debate.

\subsubsection{Fuzzballs}\label{sec:fuzz}

 The idea of firewalls and the AMPS argument motivating it are of a  general nature, not tied to string theory, or any particular account of the nature or formation of the firewall. However, there is a string theoretical proposal along the lines of a firewall -- certainly it falls into the category of ``drama at the horizon'' (though it predates the AMPS paper, e.g., \citet{Mat:05}). The example will, for string theory: (i) help illuminate how unitary evaporation might be obtained; (ii) give a concrete proposal for the interior of a black hole, in which the spacetime description may break down; (iii) so that a classical spacetime cannot be assumed, and the information `paradox' must be addressed.

\cite{Mat:09} shows that one cannot escape Hawking's argument by small quantum corrections adding a small amount of quantum `hair' to the black hole, which might account for apparently lost information. Rather, avoiding information loss requires a great deal of quantum hair -- a `fuzzball' of such hair, in fact! The work of Mathur and collaborators explores a string theoretic model of just this kind, with significant consequences for spacetime in black hole models as we shall now explain -- though not before noting that some of this work is controversial, even in the string theory community, unlike the preceding.

 In modeling a stringy black hole, one specifies mass and charge at a point of the large spatial dimensions, suitable to produce it according to the supergravity EFE (to avoid the Weyl anomaly). But the sources must ultimately be understood as an \emph{extended} stringy system, not point properties. Moreover, as we discussed, one cannot interpret the brane system as the literal source of the black hole, for it is studied with the string coupling turned off, so that the specific states counted are not gravitational at all. As a result, the model is really only valid `sufficiently far from' the sources, and does not tell us the geometry of the black hole in their vicinity -- in particular about what happens close to the classical singularity.

Of course one would like to know that, but such ignorance does not immediately cast doubt on the rest of the construction. The string system is supposed to be at a point in the large spacetime dimensions in which the horizon forms, and prima facie, it is reasonable to suppose that the size of the string system is no more than the Planck or string length, thus far from the horizon until the last stages of evaporation. Hence one expects that the derived geometry describes most of the black hole interior accurately; that string theory agrees with classical supergravity except near the singularity. However, explicit calculations of the string dynamics show that the stringy objects producing the black hole vibrate in its interior, so are not truly located at a point, but apparently extend to form a fuzzball. Thus the prima facie argument cannot be trusted, and one has to ask how large the vibrations are to discover how much of the black hole is occupied by the fuzzball. 

Studying the fuzzball in detail will reveal the detailed string state, and hence the geometry of the region that contains it -- if indeed the state of the fuzzball corresponds to a classical geometry at all. The calculation (reviewed in \cite{Mat:12}) exploits a duality between the stringy source of the black hole and a long, floppy string, to estimate the size of the vibrations, so of the fuzzball: around $(g^2\alpha'\sqrt{Q_1Q_5N}/RV)^{-1/3}$ -- which happens to be the radius of the horizon! Contrary to expectations, the fuzzball is not confined to the center, but apparently fills the black hole.

Invoking AdS-CFT duality, \cite{Mat:12} argues for the following picture (details are well beyond the scope of this paper). The fuzzball is not isotropic, as one implicitly assumes by taking it to be a point source in finding the supergravity black hole geometry; when anisotropy is taken into account the solution no longer has a horizon or singularity (\cite{lunin2002gravity}). In fact, the fuzzball causes the compact, cylindrical and toroidal, dimensions to  `cap off' at the horizon, though leaving the geometry away from the black hole essentially unchanged. Picture a cylinder smoothly tapering to a curved end; the open dimension ends where the circular dimension shrinks to a point, as there is no more space to travel into. Something similar happens around the fuzzball, although the geometry is more complicated (a `Kaluza-Klein monopole'); the compact dimensions ($S^1$ and $T^4$ in our model) apparently cap off at the horizon radius, similarly terminating any trajectory into the black hole -- a `fauxrizon', marking the end of space external to the black hole!

What is beyond this fauxrizon? The results just quoted apply to individual states of the fuzzball; from that point of view there is no `interior' strictly speaking, and no `beyond' in a spatial sense, just a nonspatial, fundamentally stringy state. However, Mathur's group has shown how approximate spatial structure might be attributed to an `interior', in terms of suitable statistical averages of states: internal space as a kind of thermodynamical property of the fuzzball. Thus if one asks, in a more operationalist spirit, what happen if you throw something through the fauxrizon, there are two possible answers. Perhaps the object `sees' the thermodynamical space in the interior and passes through; in a more fundamental description, the result of some complex interaction with the fuzzball is that the object emerges on the other side, changed to reflect an apparent passage through it. Or perhaps, objects are simply amalgamated into the fuzzball state at the fauxrizon; after all, both fuzzball and matter are ultimately just complicated compositions of the same fundamental objects of string theory. In that case, operationally there truly is no interior -- and there is plenty of `drama' at the fauxrizon!

In either case, there is no horizon to cause an information paradox, and the fuzzball models recover both $S_{BH}$, and the Hawking radiation rate. But particularly in the latter case, we have an example where it is obviously inappropriate to ascribe a classical geometry to the interior, along the lines suggested by Polchinski earlier. Clearly in this case, black holes have rather profound implications for the nature of spacetime! Moreover, Maudlin's construction again does not apply; unitarity -- and indeed information conservation -- is obtained by the details of the fuzzball dynamics.

\section{Conclusion: implications for the nature of spacetime and matter}\label{sec:lessons}

In this paper we have reviewed how black holes can be modeled in string theory. While our main focus is on questions of ontology we also briefly addressed epistemological questions, arguing that the derivation of BHT from string theory should be understood more as a consistency check, and weakly rather than as strongly confirmatory.  We emphasize, however, that the importance of the models lies in giving a successful account of the underlying states, and so providing a Boltzmannian understanding of the entropy.  

However, the main purpose of the paper is to throw some light on the ontological questions about spacetime and matter. To that end we have explicated the `standard' interpretation of classical spacetime and matter according to string theory (\S\ref{sec:strings}): both classical matter and geometry correspond to coherent states of strings in suitable excitations. Then we described its application to a black hole model  (\S\ref{sec:stringBH}), and investigated  some of the possible implications for BHT  and introduced the fuzzball proposal (\S\ref{sec:info}). In this, concluding, section we draw some further lessons for the status of spacetime in string theory.

First, the standard interpretation applies to the stringy black hole: Weyl symmetry leads to GR and classical supergravity, according to which the point charges corresponding to the brane system produce a horizon in the spacetime geometry. Alternatively, if stringy matter is in fact a fuzzball, or there is a firewall, then the spacetime description breaks down at the `horizon'; the geometry is as before outside, but the `inside' is purely stringy. Either way, classical spacetime geometry is an effective description of a multi-string coherent state (and not a fundamental, classical geometry).
  
Second, empirical significance of the derived structure -- the metric $g_{\mu\nu}$ -- comes in the first place from its role in determining scattering amplitudes: it appears in the path integrals (\ref{eq:poly}) and (\ref{eq:SUSY}) and so different values lead to different cross-sections for observed particle scattering. However, stringy astrophysical models like black holes demonstrates further significance: astronomical observations of spacetime structure are understood as low-resolution observations of fundamental stringy fields. These points show that one has to be cautious with the claim that string theory has no empirical consequences: it reproduces the predictions of GR including observable objects like black holes (and scattering amplitudes, though not yet of the standard model). What it lacks (so far) are specific \emph{novel} predictions, testable using current technologies. 
   
Third, how cogent is the standard interpretation? The most questionable point concerns the existence (at least approximately) of suitable coherent states: string theory as developed is inherently perturbative, and the possibility of such states is postulated for an unknown exact theory. That is no argument against the picture, and indeed once the basic framework of perturbative string theory is accepted, it is a small step to coherent states; but the point does emphasize how the interpretation is speculative.
     
Fourth, supposing that coherent string states exist, and that they have an effective description as coherent states of quanta, one must ask about the classical limit: as a general question about QFT, do coherent states adequately  play the role of observed classical fields? Given the importance of the assumption that they do, there is remarkably little discussion of this question in the literature\footnote{\cite{Ros:13} is a significant exception, and it is explored further in \cite{Nic:20}.}, but one question in the present case is whether graviton coherent states remain coherent long enough to model cosmological scenarios? States will retain their coherence, and classical-like behavior, only if their equations of motion are linear; so graviton coherent states will certainly lose their coherence, because of the non-linearity of the field equations. But on what time scales should we expect to see non-classical, quantum behaviour as a result? On the one hand, for a Schwarzschild black hole in the Wheeler-DeWitt framework \cite{KieLou:98} find the dispersion time to be $10^{73}\times(\mathrm{mass\ in\ solar\ masses})^3$ seconds -- a comforting 56 orders of magnitude greater than the age of the universe (and of the order of the Hawking radiation time) for a solar mass black hole! On the other, \citet[\S3.3]{Wal:12} points out that the chaotic nature of less symmetric gravitational systems can lead to a rapid loss of coherence. So matters are unclear.

Finally, we return to Bokulich and Curiel's question regarding the relation between matter and physical geometry. According to the standard interpretation the `conversion' of classical matter to geometry, and the reconversion of geometry back to matter in the form of quantum radiation, is ultimately a transition between different multi-string states. In the first case from states of strings in a matter mode to  states of strings in a graviton mode; in the latter, back from states of stringy gravitons to stringy matter quanta. Once again, with only perturbative string theory in hand one does not have a full theory of how these transitions occur. However, the mechanism (\ref{eq:rad}) provides a model for  such a transition: brane excitations decaying to stringy quanta, illustrating the dynamical fungibility of geometry and matter. Even though, as noted, the brane system is not the literal black hole interior, its correspondence with the black hole system indicates that string processes do indeed underwrite geometry-matter conversion.

\bibliographystyle{plainnat}

\bibliography{biblio2}

\begin{thebibliography}{55}
\providecommand{\natexlab}[1]{#1}
\providecommand{\url}[1]{\texttt{#1}}
\expandafter\ifx\csname urlstyle\endcsname\relax
  \providecommand{\doi}[1]{doi: #1}\else
  \providecommand{\doi}{doi: \begingroup \urlstyle{rm}\Url}\fi

\bibitem[Almheiri et~al.(2013)Almheiri, Marolf, Polchinski, and
  Sully]{AlmMar:13}
Ahmed Almheiri, Donald Marolf, Joseph Polchinski, and James Sully.
\newblock Black holes: complementarity or firewalls?
\newblock \emph{Journal of High Energy Physics}, 2013\penalty0 (2):\penalty0
  62, 2013.

\bibitem[Bekenstein(1973)]{Bek:73}
Jacob~D. Bekenstein.
\newblock Black holes and entropy.
\newblock \emph{Physical Review D}, 7\penalty0 (8):\penalty0 2333, 1973.

\bibitem[Belot et~al.(1999)Belot, Earman, and Ruetsche]{BelEar:99}
Gordon Belot, John Earman, and Laura Ruetsche.
\newblock The {H}awking information loss paradox: The anatomy of controversy.
\newblock \emph{The British journal for the philosophy of science}, 50\penalty0
  (2):\penalty0 189--229, 1999.

\bibitem[Callan and Maldacena(1996)]{ca:96}
Curtis~Gove Callan and Juan~M. Maldacena.
\newblock D-brane approach to black hole quantum mechanics.
\newblock \emph{Nuclear Physics B}, 472\penalty0 (3):\penalty0 591--608, 1996.

\bibitem[Crowther et~al.(2019)Crowther, Linnemann, and W{\"u}thrich]{Cro:19}
Karen Crowther, Niels~S. Linnemann, and Christian W{\"u}thrich.
\newblock What we cannot learn from analogue experiments.
\newblock \emph{Synthese}, \penalty0 (1-26), 2019.

\bibitem[Curiel and Bokulich(2012)]{CurBok:12}
Erik Curiel and Peter Bokulich.
\newblock Singularities and black holes.
\newblock In Edward~N. Zalta, editor, \emph{The Stanford Encyclopedia of
  Philosophy}. Metaphysics Research Lab, Stanford University, fall 2012
  edition, 2012.

\bibitem[Dardashti et~al.(2017)Dardashti, Th{\'e}bault, and
  Winsberg]{DarThe:17}
Radin Dardashti, Karim~P.Y. Th{\'e}bault, and Eric Winsberg.
\newblock Confirmation via analogue simulation: What dumb holes could tell us
  about gravity.
\newblock \emph{British Journal for the Philosophy of Science}, 68\penalty0
  (1):\penalty0 55--89, 2017.

\bibitem[Das and Mathur(2000)]{DasMat:00}
Sumit~R. Das and Samir~D. Mathur.
\newblock The quantum physics of black holes: Results from string theory.
\newblock \emph{Annual Review of Nuclear and Particle Science}, 50\penalty0
  (1):\penalty0 153--206, 2000.

\bibitem[De~Haro et~al.(2020)De~Haro, van Dongen, Visser, and
  Butterfield]{de20}
Sebastian De~Haro, Jeroen van Dongen, Manus Visser, and Jeremy Butterfield.
\newblock Conceptual analysis of black hole entropy in string theory.
\newblock \emph{Studies in History and Philosophy of Science Part B: Studies in
  History and Philosophy of Modern Physics}, 69:\penalty0 82--111, 2020.

\bibitem[Duncan(2012)]{Dun:12}
Anthony Duncan.
\newblock \emph{The conceptual framework of quantum field theory}.
\newblock Oxford University Press, 2012.

\bibitem[Green et~al.(1987)Green, Schwarz, and Witten]{GreSch:87}
Michael~B. Green, John~H. Schwarz, and Edward Witten.
\newblock \emph{Superstring Theory}, volume~I.
\newblock Cambridge University Press Cambridge, 1987.

\bibitem[Harlow(2016)]{Har:16}
Daniel Harlow.
\newblock Jerusalem lectures on black holes and quantum information.
\newblock \emph{Reviews of Modern Physics}, 88\penalty0 (1):\penalty0 015002,
  2016.

\bibitem[Hawking(1975)]{Haw:75}
Stephen~W. Hawking.
\newblock Particle creation by black holes.
\newblock \emph{Communications in mathematical physics}, 43\penalty0
  (3):\penalty0 199--220, 1975.

\bibitem[Hawking(1976)]{Haw:76}
Stephen~W. Hawking.
\newblock Breakdown of predictability in gravitational collapse.
\newblock \emph{Physical Review D}, 14\penalty0 (10):\penalty0 2460, 1976.

\bibitem[Horowitz et~al.(1996)Horowitz, Maldacena, and Strominger]{HorMal:96}
Gary~T Horowitz, Juan~M Maldacena, and Andrew Strominger.
\newblock Nonextremal black hole microstates and u-duality.
\newblock \emph{Physics Letters B}, 383\penalty0 (2):\penalty0 151--159, 1996.

\bibitem[Huggett(2015)]{Hug:15}
Nick Huggett.
\newblock Target space $\ne$ space.
\newblock \emph{Studies in history and philosophy of science part B: Studies in
  history and philosophy of modern physics}, 2015.

\bibitem[Huggett and Vistarini(2015)]{HugVis:15}
Nick Huggett and Tiziana Vistarini.
\newblock Deriving general relativity from string theory.
\newblock \emph{Philosophy of Science}, 82\penalty0 (5):\penalty0 1163--1174,
  2015.

\bibitem[Huggett and W{\"u}thrich(2013)]{HugWut:13a}
Nick Huggett and Christian W{\"u}thrich.
\newblock Emergent spacetime and empirical (in) coherence.
\newblock \emph{Studies in History and Philosophy of Science Part B: Studies in
  History and Philosophy of Modern Physics}, 44\penalty0 (3):\penalty0
  276--285, 2013.

\bibitem[Huggett and W\"uthrich(forthcoming)]{Nic:20}
Nick Huggett and Christian W\"uthrich.
\newblock \emph{Out Of Nowhere}.
\newblock Oxford University Press, forthcoming.

\bibitem[Karaca(2012)]{Kar:12}
Koray Karaca.
\newblock Kitcher's explanatory unification, kaluza-klein theories, and the
  normative aspect of higher dimensional unification in physics.
\newblock \emph{British Journal for the Philosophy of Science}, 63\penalty0
  (2):\penalty0 287--312, 2012.

\bibitem[Kiefer and Louko(1998)]{KieLou:98}
Claus Kiefer and Jorma Louko.
\newblock Hamiltonian evolution and quantization for extremal black holes.
\newblock \emph{arXiv preprint gr-qc/9809005}, 1998.

\bibitem[Lubo{\v s}(2012)]{Mot:12}
Motl Lubo{\v s}.
\newblock What is background independence and how important is it?, 2012.
\newblock URL
  \url{http://motls.blogspot.com/2012/12/what-is-background-independence-and-how.html}.

\bibitem[Lunin et~al.(2002)Lunin, Maldacena, and Maoz]{lunin2002gravity}
Oleg Lunin, Juan Maldacena, and Liat Maoz.
\newblock Gravity solutions for the d1-d5 system with angular momentum.
\newblock \emph{arXiv preprint hep-th/0212210}, 2002.

\bibitem[Mathur(2005)]{Mat:05}
Samir~D. Mathur.
\newblock The fuzzball proposal for black holes: An elementary review.
\newblock \emph{Fortschritte der Physik: Progress of Physics}, 53\penalty0
  (7-8):\penalty0 793--827, 2005.

\bibitem[Mathur(2009)]{Mat:09}
Samir~D Mathur.
\newblock The information paradox: a pedagogical introduction.
\newblock \emph{Classical and Quantum Gravity}, 26\penalty0 (22):\penalty0
  224001, 2009.

\bibitem[Mathur(2012)]{Mat:12}
Samir~D. Mathur.
\newblock Black holes and beyond.
\newblock \emph{Annals of Physics}, 327\penalty0 (11):\penalty0 2760--2793,
  2012.

\bibitem[Matsubara(2013)]{Mat:13}
Keizo Matsubara.
\newblock Realism, underdetermination and string theory dualities.
\newblock \emph{Synthese}, 190\penalty0 (3):\penalty0 471--489, 2013.

\bibitem[Maudlin(2017)]{Mau:17}
Tim Maudlin.
\newblock (information) paradox lost.
\newblock \emph{arXiv preprint arXiv:1705.03541}, 2017.

\bibitem[Misner et~al.(1973)Misner, Thorne, and Wheeler]{MisTho:73}
Charles~W. Misner, Kip~S. Thorne, and John~Archibald Wheeler.
\newblock \emph{Gravitation}.
\newblock Freeman, 1973.

\bibitem[Polchinski(1995)]{pol:95}
Joseph Polchinski.
\newblock Dirichlet branes and ramond-ramond charges.
\newblock \emph{Physical Review Letters}, 75\penalty0 (26):\penalty0 4724,
  1995.

\bibitem[Polchinski(1998)]{Pol:03}
Joseph Polchinski.
\newblock \emph{String theory}.
\newblock Cambridge university press, 1998.

\bibitem[Polchinski(2017)]{Pol:17}
Joseph Polchinski.
\newblock The black hole information problem.
\newblock In \emph{New Frontiers in Fields and Strings: TASI 2015 Proceedings
  of the 2015 Theoretical Advanced Study Institute in Elementary Particle
  Physics}, pages 353--397. World Scientific, 2017.

\bibitem[Read(2019)]{Rea:19}
James Read.
\newblock On miracles and spacetime.
\newblock \emph{Studies in History and Philosophy of Science Part B: Studies in
  History and Philosophy of Modern Physics}, 65:\penalty0 103 -- 111, 2019.

\bibitem[Rosaler(2013)]{Ros:13}
Joshua~S. Rosaler.
\newblock \emph{Inter-theory relations in physics: case studies from quantum
  mechanics and quantum field theory}.
\newblock PhD thesis, University of Oxford, 2013.
\newblock URL
  \url{https://ora.ox.ac.uk/objects/uuid:1fc6c67d-8c8e-4e92-a9ee-41eeae80e145}.

\bibitem[Salimkhani(2018)]{Sal:18}
Kian Salimkhani.
\newblock Quantum gravity: A dogma of unification?
\newblock In Alexander Christian, David Hommen, Nina Retzlaff, and Gerhard
  Schurz, editors, \emph{Philosophy of Science Between the Natural Sciences,
  the Social Sciences, and the Humanities}, volume~9 of \emph{European Studies
  in Philosophy of Science}, pages 23--41. Springer International Publishing,
  2018.
\newblock ISBN 9783319725772.

\bibitem[Strominger and Vafa(1996)]{StrVaf:96}
Andrew Strominger and Cumrun Vafa.
\newblock Microscopic origin of the bekenstein-hawking entropy.
\newblock \emph{Physics Letters B}, 379\penalty0 (1-4):\penalty0 99--104, 1996.

\bibitem[Susskind(2006)]{Sus:06}
Leonard Susskind.
\newblock The paradox of quantum black holes.
\newblock \emph{Nature Physics}, 2\penalty0 (10):\penalty0 665, 2006.

\bibitem[Susskind(2012{\natexlab{a}})]{Sus:12}
Leonard Susskind.
\newblock Complementarity and firewalls.
\newblock Technical report, 2012{\natexlab{a}}.

\bibitem[Susskind(2012{\natexlab{b}})]{Sus:12a}
Leonard Susskind.
\newblock Singularities, firewalls, and complementarity.
\newblock \emph{arXiv preprint arXiv:1208.3445}, 2012{\natexlab{b}}.

\bibitem[Susskind(2012{\natexlab{c}})]{Sus:12b}
Leonard Susskind.
\newblock The transfer of entanglement: the case for firewalls.
\newblock \emph{arXiv preprint arXiv:1210.2098}, 2012{\natexlab{c}}.

\bibitem[Susskind and Lindesay(2005)]{SusLin:05}
Leonard Susskind and James Lindesay.
\newblock \emph{An introduction to black holes, information and the string
  theory revolution}.
\newblock World Scientific, 2005.

\bibitem[Susskind et~al.(1993)Susskind, Thorlacius, and Uglum]{SusTho:93}
Leonard Susskind, Larus Thorlacius, and John Uglum.
\newblock The stretched horizon and black hole complementarity.
\newblock \emph{Physical Review D}, 48\penalty0 (8):\penalty0 3743, 1993.

\bibitem[Taylor(2009)]{Tay:09}
Washington Taylor.
\newblock String field theory.
\newblock In Daniele Oriti, editor, \emph{Approaches to Quantum Gravity: Toward
  a New Understanding of Space, Time and Matter}, pages 210--28. Cambridge
  University Press, 2009.

\bibitem[van Dongen and de~Haro(2004)]{van:04}
Jeroen van Dongen and Sebastian de~Haro.
\newblock On black hole complementarity.
\newblock \emph{Studies in History and Philosophy of Science Part B: Studies in
  History and Philosophy of Modern Physics}, 35\penalty0 (3):\penalty0
  509--525, 2004.

\bibitem[van Dongen et~al.(2020)van Dongen, De~Haro, Visser, and
  Butterfield]{DonDe-:20}
Jeroen van Dongen, Sebastian De~Haro, Manus Visser, and Jeremy Butterfield.
\newblock Emergence and correspondence for string theory black holes.
\newblock \emph{Studies in History and Philosophy of Science Part B: Studies in
  History and Philosophy of Modern Physics}, 69:\penalty0 112--127, 2020.

\bibitem[Vistarini(2019)]{Vis:19}
Tiziana Vistarini.
\newblock \emph{The Emergence of Spacetime in String Theory}.
\newblock Routledge, 2019.

\bibitem[Wadia(2001)]{Wad:01}
Spenta~R. Wadia.
\newblock A microscopic theory of black holes in string theory: Thermodynamics
  and hawking radiation.
\newblock \emph{CURRENT SCIENCE-BANGALORE-}, 81\penalty0 (12):\penalty0
  1591--1597, 2001.

\bibitem[Wald(1994)]{Wal:94}
Robert~M. Wald.
\newblock \emph{Quantum field theory in curved spacetime and black hole
  thermodynamics}.
\newblock University of Chicago press, 1994.

\bibitem[Wallace(2012)]{Wal:12}
David Wallace.
\newblock \emph{The Emergent Multiverse: Quantum Theory According to the
  Everett Interpretation}.
\newblock Oxford University Press, Oxford, 2012.

\bibitem[Wallace(2018)]{Wal:18}
David Wallace.
\newblock The case for black hole thermodynamics part i: Phenomenological
  thermodynamics.
\newblock \emph{Studies in History and Philosophy of Science Part B: Studies in
  History and Philosophy of Modern Physics}, 64:\penalty0 52--67, 2018.

\bibitem[Wallace(2019)]{Wal:19}
David Wallace.
\newblock The case for black hole thermodynamics part ii: statistical
  mechanics.
\newblock \emph{Studies in History and Philosophy of Science Part B: Studies in
  History and Philosophy of Modern Physics}, 66:\penalty0 103--117, 2019.

\bibitem[Wallace(2020)]{Wal:20}
David Wallace.
\newblock Why black hole information loss is paradoxical.
\newblock In Nick Huggett, Keizo Matsubara, and Christian W{\"u}thrich,
  editors, \emph{Beyond Spacetime: The Foundations of Quantum Gravity},
  chapter~10, pages 209--236. Cambridge University Press, 2020.

\bibitem[Witten(1996)]{wit96}
Edward Witten.
\newblock Reflections on the fate of spacetime.
\newblock \emph{Physics Today}, pages 24--30, April 1996.

\bibitem[W\"uthrich(2017)]{Wut:17}
Christian W\"uthrich.
\newblock Are black holes about information?, 2017.

\bibitem[Zwiebach(2004)]{Zwi:04}
Barton Zwiebach.
\newblock \emph{A first course in string theory}.
\newblock Cambridge university press, 2004.

\end{thebibliography}

\end{document}